\documentclass{article}
\usepackage{psfig}

\newcommand{\ket}[1]{\left | \, #1 \right \rangle}
\newcommand{\sket}[1]{\big | \, #1 \big \rangle}

\newcommand{\beq}{\begin{equation}}
\newcommand{\eeq}{\end{equation}}

\begin{document}

\title{ The Hidden Subgroup Problem and
Eigenvalue Estimation on a Quantum Computer}

\author{
Michele Mosca\thanks{Clarendon Laboratory, Parks Road, Oxford, OX1
3PU, U.K. and Mathematical Institute, 24-29 St. Giles', Oxford, OX1
3LB, U.K.} \and Artur Ekert\thanks{Clarendon Laboratory, Parks
Road, Oxford, OX1 3PU, U.K.}}

\date{May 1998}

\maketitle

\begin{abstract}
A quantum computer can efficiently find the order of an element in
a group, factors of composite integers, discrete logarithms,
stabilisers in Abelian groups,
and
\emph{hidden}
or
\emph{unknown}
subgroups of Abelian groups.
It is already known how to phrase the first
four problems as the estimation of eigenvalues of
certain unitary operators.  Here we show how the solution to
the more general Abelian
\emph{hidden subgroup problem} can also
be described and analysed as such.
We then point out how
certain instances of these problems can be solved with only
one control qubit, or \emph{flying qubits},
instead of entire registers of control
qubits.

\end{abstract}

\section{Introduction}

Shor's approach to factoring \cite{Shor}, (by finding the order of
elements in
the multiplicative group of integers
mod $N$, referred to as ${\mathbf{Z}}_N^{*}$)
is to extract the period
in a superposition by applying a Fourier transform.
Another approach, based on Kitaev's technique \cite{Kitaev},
is to estimate
an eigenvalue of a certain unitary operator.
The difference between the two
analyses is that the first one
considers (or even 'measures' or 'observes')
the \emph{target} or \emph{output} register in
the standard computational basis, while the analysis we
detail here
considers the target register in a basis containing
eigenvectors of unitary operators related to the function $f$.
The actual network of quantum gates, as
highlighted in \cite{CEMM}, is
the same for both algorithms;
it is helpful to understand both approaches.
In some cases, which we discuss in Sect. \ref{control.bit},
this approach suggests implementations which do
not require a register of control qubits.
A more general formulation of the order-finding problem
as well as the discrete logarithm problem,
and the Abelian stabiliser problem, is
the \emph{hidden subgroup problem} (or the \emph{unknown}
subgroup problem \cite{Hoyer}).
In the case that
$G$ is presented as the product
of a finite number of cyclic groups (so
$G$ is finitely generated and Abelian),
all of these problems are solved by the familiar
sequence of a Fourier transform, a function application,
and an inverse Fourier transform.
In this paper we describe how this more general problem
can also be viewed and analysed as an estimation of eigenvalues
of unitary operators.

\section{The Hidden Subgroup Problem} \label{unknown.subgroup}

\begin{figure}
\psfig{figure=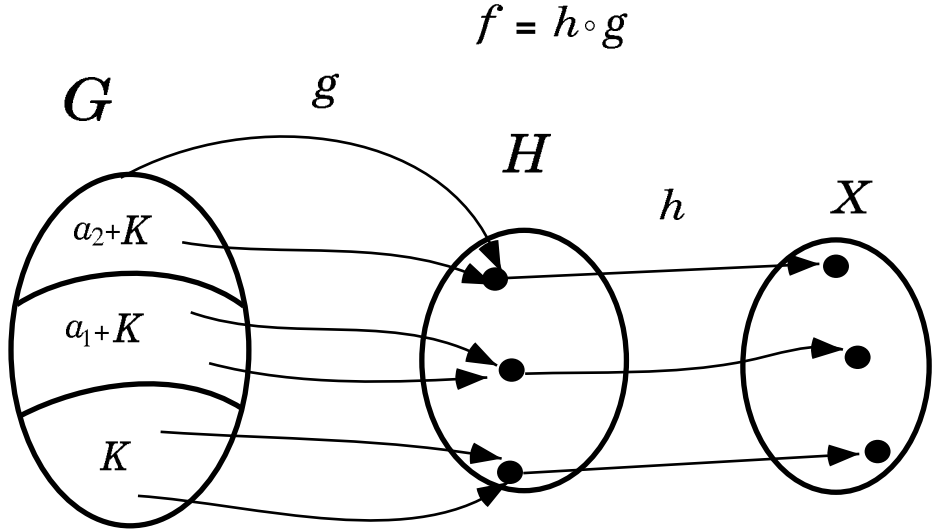}
\caption[ ]{ The function $f$ can be viewed
as the composition of a homomorphism $g$ to
a group $H$, and some 1-to-1 mapping $h$
to the set $X$.
Our hidden subgroup $K$ will be the
kernel of $g$, and $H$ is isomorphic to $G/K$.
}
\label{draw_f}
\end{figure}

Let $f$ be a function from a finitely generated group $G$ to a
finite set $X$
such that $f$ is constant on the cosets of a subgroup $K$ (of finite
index, since $X$ is finite),
and distinct on each coset.
The hidden subgroup problem is to find $K$ (that is, a generating
set for $K$), given
a way of computing $f$.
When $K$ is normal in $G$,
we could in fact decompose $f$ as $h \circ g$, where
$g$ is a homomorphism from $G$ to some finite group $H$,
and $h$ is some 1-to-1 mapping from $H$ to the set $X$.
In this case, $K$ corresponds to the kernel of $g$
and $H$ is isomorphic to $G/K$.
We will occasionally refer to this decomposition,
which we illustrate in Fig. \ref{draw_f}.
Define the \emph{input size}, $n$,
to be of order $\log_2 [G:K]$.
We will count the
number of operations, or the \emph{running time},
in terms of $n$.
An algorithm is considered \emph{efficient}
if its running time is polynomial in the input size.
By \emph{elementary quantum operations},
we are referring to a finite set of quantum logic
gates which
allow us to approximate any unitary operation.
See \cite{BBCDMSSSW} for a discussion and further
references.
Our running times will always refer to expected
running times, unless explicitly stated otherwise.
By \emph{expected running time} we are referring to
the expected number of operations for \emph{any}
input (and not just an average of the
expected running times over all inputs).

We should be clear about what it means to have
a finitely generated group $G$,
and to be able to compute the function $f$.
This is difficult without losing some generality or
being dry and technical, or both.
The algorithms we describe only apply for groups
$G$ which are represented as finite tuples of integers
corresponding to the direct product of cyclic groups (consequently,
$G$ is finitely generated and Abelian).
Conversely,
for any finitely generated Abelian $G$, there is a temptation
to point out that $G$ is isomorphic to such a direct product of
cyclic groups, and \emph{assume}
that we can easily access this product
structure.  This is not always the case, even in cases of
practical interest.
For example, ${\mathbf{Z}}_N^{*}$, the multiplicative group of
integers modulo $N$ for some large integer $N$, which
is Abelian of order $\phi(N)$ (the Euler $\phi$-function)
and thus isomorphic to a product of cyclic groups of prime power
order.
We will not necessarily know $\phi(N)$ or have
a factorisation of it along with a set of generators
for ${\mathbf{Z}}_N^{*}$.
However, in light of the quantum algorithms described in this
paper, we could efficiently find such an isomorphism,
thereby increasing the number of finitely generated
Abelian groups which can be efficiently expressed
in a manner which allows us to employ these algorithms.
We will however leave further discussion of these
details to another note \cite{EM.groups}.
When we talk about computing $f$, we assume that we have
some unitary operation $U_f$ which takes us from state
$\ket{\mathbf{x}}\ket{\mathbf{0}}$
to $\ket{\mathbf{x}}\ket{f({\mathbf{x}})}$.
It could, for example, take
$\ket{\mathbf{x}}\ket{\mathbf{y}}$
to $\ket{\mathbf{x}}\ket{{\mathbf{y}}+f({\mathbf{x}})}$,
where $+$ denotes an appropriate group operation,
such as addition modulo $N$ when the second register is
used to represent the integers modulo $N$.

Various cases of the hidden subgroup problem are described in \cite{Simon},
\cite{Shor}, \cite{Kitaev}, \cite{Boneh-Lipton},
\cite{Grigoriev}, \cite{Jozsa}, \cite{CEMM}, and \cite{Hoyer}.
We note that \cite{Boneh-Lipton} also covers the case
that $f$ is not necessarily distinct on each coset (that is,
$h$ is not 1-to-1), and
this is discussed in the appendix.
Finding the order $r$ of an element in a group $H$
of unknown size, or the period $r$
of a function $f$, is a special case where $G = {\mathbf{Z}}$
and $K = r{\mathbf{Z}}$.   For any generator $\mathbf{e_j}$ of
a finitely generated $G$,
we can use the
algorithm in Sect. \ref{period.find}
to find an integer $k$ such that $f(k {\mathbf{e_j}}) = f({\mathbf{0}})$,
so that $k{\mathbf{e_j}} \in K$.  We find this
$k$ with $O(n)$ applications of
$f$ and $O(n^2)$ other elementary quantum operations.
We can then
assume that $\mathbf{e_j}$
is of order $k$ (that is, factor $\langle k {\mathbf{e_j}} \rangle$
out of $G$), and in general assume that $G$ is a finite group.

We give a few examples.

{\textbf{Deutsch's Problem}}: Consider a function $f$ mapping
${\mathbf{Z}}_2 = \{0,1\}$ to $\{0,1\}$.
Then $f(x) = f(y)$ if and only if $x-y \in K$, where
where $K$ is either $\{0\}$ or
${\mathbf{Z}}_2 = \{0,1\}$.  If $K$ is $\{0\}$,
then $f$ is $1-to-1$  (or \emph{balanced}), and
if $K$ is
${\mathbf{Z}}_2$ then $f$ is constant.
\cite{Deutsch} \cite{CEMM}
\vspace{3mm}

{\textbf{Simon's Problem}}:
Consider a function $f$ from ${{\mathbf{Z}}_2}^l$ to some set $X$
with the property that $f(x) = f(y)$ if and only
if $x - y \in \{ \mathbf{0}, \mathbf{s} \}$
for some string $\mathbf{s}$ of length $l$.
Here $K = \{ \mathbf{0}, \mathbf{s} \}$ is the  hidden
subgroup of ${{\mathbf{Z}}_2}^l$.
Simon \cite{Simon} presents an efficient algorithm
for solving this problem, and the solution to the
hidden subgroup problem in the Abelian case is a generalisation.
\vspace{3mm}

{\textbf{Discrete Logarithms}}:
Let $G$ be the
group ${\mathbf{Z}}_r \times {\mathbf{Z}}_r$ where ${\mathbf{Z}}_r$
is the additive group of integers modulo $r$.
Let the set $X$ be the subgroup generated by some element
$a$ of a group $H$, with $a^r = 1$.
For example, $H = {\mathbf{F}}_q^{*}$,
the multiplicative group of the field of order $q$, where
$r = q-1$.
Let $a,b \in G$, and suppose $b = a^m$.
Define $f$ to map $(x,y)$ to $a^x b^y$.
Here the hidden subgroup of $G$ is
$K = \{ (k, -km) | k = 0,1, \ldots, r-1 \} = \langle (1, -m) \rangle$,
the subgroup generated by $(1,-m)$.
Finding this hidden subgroup will give us the logarithm of
$b$ to the base $a$.
The security of the U.S. Digital Signature Algorithm is based on
the computational difficulty of this problem in ${\mathbf{F}}_q^{*}$
(see \cite{MOV} for details and references).
Here the input size is $n = \lceil \log_2 r \rceil$.
Shor's algorithm \cite{Shor} was the first to solve
this problem efficiently.
In this case, $f$ is also a homomorphism
which can make implementations more simple
as described in Sect. \ref{control.bit}.
\vspace{3mm}

{\textbf{Self-Shift-Equivalent Polynomials}}:
Given a polynomial $P$ in $l$ variables
$X_1$, $X_2$, $\ldots, X_l$ over
${\mathbf{F}}_q$, the function $f$ which maps
$(a_1, a_2, \ldots, a_l) \in {\mathbf{F}}_q^l$
to $P(X_1 - a_1, X_2 - a_2, \ldots, X_l - a_l)$
is constant on cosets of a subgroup $K$
of ${{\mathbf{F}}_q^l}$.  This subgroup $K$ is the set of
self-shift-equivalences of the polynomial $P$.
Grigoriev \cite{Grigoriev} shows how to compute
this subgroup.
He also shows, in the case that $q$ has characteristic $2$,
how to decide if two polynomials $P_1$ and $P_2$
are shift-equivalent, and to generate the set of
elements $(a_1, a_2, \ldots, a_l)$ such that
$P_1(X_1 -a_1, X_2 - a_2, \ldots, X_l - a_l) = P_2(X_1, X_2, \ldots , X_l)$.
The input size $n$ is at most $l \log_2 q$.
\vspace{3mm}

{\textbf{Abelian Stabiliser Problem}}:
Let
$G$ be any group
acting on a finite set $X$.
That is, each element of $G$ acts as a map from $X$ to $X$,
in such a way that for any two elements $a,b \in G$,
$a(b(x)) = (ab)(x)$ for all $x \in X$.
For a particular element $x$ of $X$, the set of elements
which fix $x$ (that is, the elements $a \in G$ such that
$a(x) = x$), form a subgroup.  This subgroup is called
the stabiliser of $x$ in $G$, denoted $St_G(x)$.
Let $f_x$ denote the function from $G$ to $X$ which maps
$g \in G$ to $g(x)$.  The hidden subgroup
corresponding to $f_x$ is $K = St_G(x)$.
The finitely generated
Abelian case of this problem was solved by Kitaev \cite{Kitaev}, and
includes finding orders and discrete logarithms as special
cases.

\section{Phase Estimation and the Quantum Fourier Transform}

In this section, we review the relationship between
phase estimation and the quantum Fourier transform which
was highlighted in \cite{CEMM}.

The quantum Fourier transform for the cyclic group of order
$N$, $F_{N}$, maps
\[ \ket{a} \rightarrow
{1 \over \sqrt{N}} \sum_{x=0}^{N-1} e^{2 \pi i ax/N} \ket{x}.\]
So $F_N^{-1}$ maps
\[ {1 \over \sqrt{N}} \sum_{x=0}^{N-1} e^{2 \pi i ax/N} \ket{x}
\rightarrow \ket{a} .\]
More generally, for any $\phi$, $0 \leq \phi < 1$,
 $F_N^{-1}$ maps
\begin{equation}
{1 \over \sqrt{N}} \sum_{x=0}^{N-1} e^{2 \pi i {\phi x}} \ket{x}
\rightarrow \sum_{x=0}^{N-1} \alpha_{\phi,x} \ket{x}
\label{approx}
\end{equation}
where the amplitudes $\alpha_{\phi,x}$ are concentrated near
values of $x$ such that $x/N$ are good estimates
of $\phi$.  The closest estimate of $\phi$ will have amplitude
at least $4/\pi^2$.  The probability that
$x/N$ will be within $k/N$ of $\phi$ is
at least $1 - 1/(2k-1)$.  See \cite{CEMM} for details
in the case that $N$ is a power of $2$; the same proof
works for any $N$.  Thus to estimate $\phi$ such that, with
probability at least $1-\epsilon$, the error is less than
 $1/M$, we should use a control register
containing values from $0$ to $N-1$ and apply
 $F_{N}^{-1}$ for any $N \geq M(1/\epsilon +  1)/2$.
For example, if we desire an error of at most $1/2^n$
with probability at least $1 - 1/2^m$ we could
use $N = 2^{n + m}$.  In practice, it will be best
to use the $N$ that corresponds to the group that
is easiest to represent and work with in the
particular physical realisation of the quantum computer at hand.
We expect that this $N$ will be a power of two.

For convenience, we will omit normalising factors in the
remainder of this paper.
It will also be convenient to have a compact notation for
the state on the right hand side of (\ref{approx})
which we consider to be a good estimator for $\ket{\phi}$.
So let us refer to this state as $\sket{\widetilde{\phi}}_N$
or just $\sket{\widetilde{\phi}}$ if the value of $N$ is
understood.  Lastly, we will use $\mbox{exp}(x)$
to denote $e^x$.

\section{The Algorithm}

To restrict attention from finitely generated groups $G$ to
finite groups we need to know
how to solve the cyclic case (just one generator),
that is, to find the period of a function from $\mathbf{Z}$
to the set $X$.
We will first describe how to find the order of an element $a$
in a group $H$, or equivalently, the period of
the function $f: t \rightarrow a^t$, as Shor \cite{Shor} did for
the group $H = {\mathbf{Z}}_N^{*}$, the multiplicative group of integers
modulo $N$.
We will then show how to generalise it to find the period of any
function $f: {\mathbf{Z}} \rightarrow X$.  If $f$ were
a homomorphism (so $h$ is an isomorphism of $H$, when $f$ is
decomposed as $f = h \circ g$),
we would just be finding the order of $f(1)$ in $H$.  The difference
is that we are showing how to deal with a non-trivial $h$
which hides the homomorphism structure.
The details will also help explain how to find  hidden
subgroups of finite Abelian groups.

\subsection{Finding Orders in Groups}

We have an element $a$ from a group $H$
and we wish to find the smallest positive
integer $r$ such that $a^r = 1$.
The group $H$ is not necessarily
Abelian; all that matters is that the subgroup
generated by $a$ is Abelian, and this is always true.
The idea is to create an operator $U_a$ which
corresponds to multiplication by $a$  (so it maps
$\ket{y}$ to $\ket{ay}$). Since $a^r = 1$, then
$U_a^r = I$, the identity operator.  Hence
the eigenvalues of $U_a$ are $r$th roots of
unity, $exp(2 \pi i k /r)$, $k = 0,1,\ldots, r-1$.
By estimating a random eigenvalue of $U_a$, with
accuracy $1/2r^2$, we can determine the fraction
$k/r$.  The denominator (with the fraction in lowest terms)
will be a factor of $r$.
We thus seek to estimate an eigenvalue of $U_a$;
note that $U_a^r = U_{a^r}$.

For any integer $x$
define $U_{a^x}$ to be the operator that
maps $\ket{y}$ to $\ket{a^x y}$.
Define $U_{a^{\mbox{x}}}$ to
be the operator which maps $\ket{x}\ket{y}$ to
$\ket{x} U_{a^x}\ket{y} = \ket{x} \ket{a^x y}$.
Note that
$U_{a^{\mbox{x}}}$ acts on two registers and
$\mbox{x}$ is a variable which takes on the value in
the first register,
while
$U_{a^x}$ acts on one register and $x$ is fixed.
Consider the eigenvectors
\begin{equation}
\ket{\Psi_k} =
\sum_{t = 0}^{r-1} \mbox{exp}(-2\pi i kt/r)\ket{a^t} , k=0,1,\ldots,r-1,
\label{Psi_def_a}
\end{equation} of $U_{a^x}$
and respective eigenvalues
$\mbox{exp}(2 \pi i kx/r)$ .
If we start with the superposition
\[ \sum_{x = 0}^{2^l-1} \ket{x} \ket{\Psi_k} \]
and then apply $U_{a^{\mbox{x}}}$ we get
\[ \sum_{x = 0}^{2^l-1} \mbox{exp}(2\pi i kx/r ) \ket{x} \ket{\Psi_k} .\]
As discussed in the previous section, applying $F_{2^l}^{-1}$
to the first register gives $\sket{\widetilde{k/r}} \ket{\Psi_k}$
and thus a good estimate of $k/r$.

We will not typically have $\ket{\Psi_k}$ but we do
know that $\ket{1} =  \sum_{k=0}^{r} \ket{\Psi_k}$.
Therefore we can start with
\begin{equation}
\ket{0}\ket{1} =\ket{0} \sum_{k=0}^{r} \ket{\Psi_k}
=\sum_{k=0}^{r}\ket{0}\ket{\Psi_k}
\label{order.one}
\end{equation}
and then
apply $F_{2^l}$ to the first register to produce
\begin{equation}
\sum_{k=0}^{r-1}
\left( \sum_{x=0}^{2^l-1} \ket{x} \right) \ket{\Psi_k}.
\label{order.two}
\end{equation}
We then apply $U_{a^{\mbox{x}}}$  to get
\begin{equation}
\sum_{k=0}^{r-1}
\left( \sum_{x=0}^{2^l-1}
\mbox{exp}(2 \pi i kx/r) \ket{x} \right) \ket{\Psi_k}
\label{order.three}
\end{equation}
followed by $F_{2^l}^{-1}$ on the control register  to yield
\begin{equation}
\sum_{k=0}^{r-1} \sket{\widetilde{k/r}} \ket{\Psi_k} .
\label{order.four}
\end{equation}

Observing the first register will give an estimate
of $k/r$ for an integer $k$ chosen uniformly at random
from the set $\{ 0,1, \ldots, r-1 \}$.
As shown in \cite{Shor},
we choose $l > 2 \log_2 r$, and use the continued
fractions algorithm to find the fraction $k/r$.
Of course, we do not know $r$, so we must either
use an $l$ we know will be larger than $2 \log_2 r$,
such as $2 \log_2 N$ in the  case that $H$ is $\mathbf{Z}_N^{*}$.
(Alternatively, we could guess a lower bound for
$r$, and if the algorithm fails, subsequently double
the guess and repeat.)
We then repeat $O(1)$ times
to find $r$.  This algorithm thus uses
$O(1)$ exponentiations, or $O(n)$ group
multiplications, and $O(n^2)$ elementary quantum operations
to do the Fourier transforms.

We can factor the integer $N$ by finding orders of
elements in ${\mathbf{Z}}_N^{*}$.
This uses only
$O(n^3)$ or $\mbox{exp}(c\log n)$
 elementary quantum operations, for $c = 3 + o(1)$
(or $c = 2 + o(1)$ if we use fast Fourier transform
techniques).  Other deterministic factoring methods
will factor $N$ in $O(\sqrt{N})$ or $\mbox{exp}(cn)$ steps,
where $c = 1/2 + o(1)$.
The best known rigorous probabilistic classical
algorithm (using index calculus methods) \cite{Lenstra.Pomerance}
uses $\mbox{exp}(c(n \log{n})^{1/2})$ elementary
classical operations,
$c = 1 + o(1)$.
There is also an algorithm with a heuristic expected
running time of
$\mbox{exp}(c(n^{1/3}(\log{n})^{2/3})$ elementary  classical
operations (see \cite{MOV} for an overview and references) for
$c = 1.902 + o(1)$.
Thus, in terms of elementary operations,
a quantum computer provides a drastic
improvement over known classical methods to factor integers.

\subsection{Finding the Period of a Function} \label{period.find}

The above algorithm, as pointed out in \cite{Boneh-Lipton},
can be applied to a more general setting.
Replace the mapping from $t$ to $a^t$ with any function
$f$ from the integers to some finite set $X$.
Define $U_{f(x)}$ to be an operator that
maps $f(y)$ to  $f(y+x)$.
This is a generalisation of $U_{a^x}$
except it does not matter how it is defined
on values not in the range of $f$, as long as it is
unitary.
Define $U_{f(\mbox{x})}$ to
be an operator which maps $\ket{x}\ket{f(y)}$ to
$\ket{x} U_{f(x)}\ket{f(y)} = \ket{x} \ket{f(y+x)}$.

The following are eigenvectors of $U_{f(x)}$:
\begin{equation}
 \ket{\Psi_k}
 = \sum_{t = 0}^{r-1} \mbox{exp}(-2\pi i kt/r)\ket{f(t)} , k=0,1, \ldots, r-1,
\label{Psi_def_gen}
\end{equation}
with respective eigenvalues
$ \mbox{exp}(2 \pi i kx/r)$.
As in (\ref{order.one}), we can start with
\[ \ket{0} \ket{f(0)}  = \sum_{k=0}^{r-1}\ket{0} \ket{\Psi_k} \]
except with our new, more general, definition of $\ket{\Psi_k}$.
We apply $F_{2^n}$ to the first register to produce
(\ref{order.two}),
and then apply $U_{f(\mbox{x})}$ to produce  (\ref{order.three}),
followed by $F_{2^n}^{-1}$ to get (\ref{order.four}).
Observing the first register will give an estimate
of $k/r$ for an integer $k$ chosen uniformly at random,
and the same analysis as in the previous section applies
to find $r$.

One important issue is how to compute $U_{f(\mbox{x})}$
only knowing how to
compute $f$.
Note that from (\ref{order.two})
to
(\ref{order.three}) (using the modified definition of $\ket{\Psi_k}$)
we simply go from
\begin{equation}
 \sum_{x=0}^{2^n-1} \ket{x} \ket{f(0)}
=
\sum_{x=0}^{2^n-1} \left( \sum_{k=0}^{r-1}\ket{x} \ket{\Psi_k} \right)
\end{equation}
to
\begin{equation}
\sum_{x=0}^{2^n-1} \ket{x} \ket{f(x)} =
\sum_{x=0}^{2^n-1} \ket{x}
\left(
\sum_{k=0}^{r-1}
\mbox{exp}(2 \pi i xk/r)
\ket{\Psi_k}
\right)
\label{main.equality}
\end{equation}
which could be accomplished by applying $U_f$, which we do have, to
the starting state
\[ \sum_{x=0}^{2^n-1} \ket{x} \ket{0} .\]
Thus even if we do not know how to explicitly compute
the operators $U_{f(x)}$, any operator $U_f$ which computes
the function $f$ will give us the state (\ref{main.equality}).
This state permits us to estimate an eigenvalue of $U_{f(x)}$
which lets us find the period of the function $f$
with just $O(1)$ applications of the operator $U_f$
and $O(n^2)$  other elementary operations.
The equality in (\ref{main.equality}) is the key to the
equivalence between the two approaches
to these quantum algorithms.  On the left
hand side is the original approach (\cite{Simon}, \cite{Shor},
\cite{Boneh-Lipton})
which considers the target register in the standard
computational basis.  We can analyse the Fourier
transform of the preimages of these basis states,
which is less easy when the Fourier transforms do not
exactly correspond to the group $G$.
On the right hand side of
(\ref{main.equality}) we consider the target
register in a basis containing the eigenvectors of the unitary operators
which we apply to it (as done in \cite{Kitaev} and \cite{CEMM}, for
example),
and this gives us (\ref{order.three}),
from which it is easy to see and analyse the effect of the
inverse Fourier transform even when it does not perfectly match
the size of $G$.

\subsection{Finding Hidden Subgroups}

As discussed in Sect. \ref{unknown.subgroup},
any finite Abelian group $G$ is the product of
cyclic groups.
In light of the order-finding algorithm, which also
permits us to factor, we can assume
that the group $G$ is represented as a product
of cyclic groups of prime power order.
Further, for any product of two groups $G_p$
and $G_q$ whose orders are coprime, any subgroup
$K$ of $G_p \times G_q$ must be equal to
$K_p \times K_q$ from some subgroups $K_p$ and $K_q$
of $G_p$ and $G_q$ respectively.
We can therefore consider our function $f$ separately on
$G_p$  and $G_q$ and determine $K_p$ and $K_q$ separately.
Thus we can further restrict ourselves to groups $G$ of prime power order.
This not only simplifies any analysis, it could reduce the
size of quantum control registers necessary in any implementation
of these algorithms.

Let us thus assume that
$G ={\mathbf{Z}}_{p^{m_1}} \times {\mathbf{Z}}_{p^{m_2}} \times \cdots
\times {\mathbf{Z}}_{p^{m_l}}$ for some prime $p$ and positive integers
$m_1 \leq m_2 \leq \cdots \leq m_l = m$. The `promise' is that $f$
is constant on cosets of a subgroup $K$, and distinct on each
coset. The hidden subgroup $K$ is $\{ {\mathbf{k}} = (k_1, k_2,
\ldots, k_l) | f({\mathbf{x}}) = f({\mathbf{x}} + {\mathbf{k}})
\hspace{2mm} \mbox{ for all } \hspace{2mm} {\mathbf{x}} \in G \}$. In
practice, this will usually be a consequence of the nature of $f$,
as in the case of discrete logarithms where $f(x_1,x_2) = a^{x_1}
b^{x_2}$, or whenever $f$ is constructed as $h \circ g$ for some
homomorphism $g$ from $G$ to some finite group $H$, and a  1-to-1
mapping $h$ from $H$ to the set $X$.

Let $U_f$ be an operator which maps
$\ket{\mathbf{x}} \ket{\mathbf{0}}$ to
$\ket{\mathbf{x}} \ket{f(\mathbf{x})}$.
Define ${\mathbf{e_1}} = (1,0, \ldots, 0)$,
${\mathbf{e_2}} = (0,1,0,\ldots, 0)$, and so on.
Let us also consider an operator related to $U_f$, $U_{f(\mbox{x}{\mathbf{e_j}})}$,
which maps $\ket{x}\ket{f(\mathbf{y})}$ to
$\ket{x}U_{f(x{\mathbf{e_j}})}\ket{f(\mathbf{y})}
= \ket{x}\ket{f({\mathbf{y}}+x{\mathbf{e_j}})}$.
In the case of Simon's Problem, the operator
$U_{f(\mbox{x}(0,1,0))}$ maps $\ket{1}\ket{f(y_1, y_2, y_3)}$ to
$\ket{1}U_{f(0,1,0)}\ket{f(y_1, y_2 , y_3))}$ $=\ket{1}\ket{f(y_1,
y_2+1 , y_3)}$ and does nothing to $\ket{0}\ket{f(y_1, y_2, y_3)}$.

For each ${\mathbf{t}} = (t_1, t_2, \ldots, t_l)$, $0 \leq t_j < p^{m_j}$,
satisfying
\begin{equation}
 \sum_{j = 1}^l {p^{m - m_j}h_j t_j } = 0 \mbox{ mod } p^{m} \hspace{2mm}
\mbox{ for all }
{\mathbf{h}} \in K
\label{orth}
\end{equation}
define
\begin{equation}
 \ket{\Psi_{\mathbf{t}}} =
\sum_{\mathbf{a} \in G/K}
\mbox{exp}\left({-2 \pi i \over p^m} \sum_{j=1}^l {p^{m-m_j} t_j a_j }\right)
\ket{f(\mathbf{a})} .
\end{equation}
We are summing over a set of representatives of the cosets
of $K$ modulo $G$,
and by condition (\ref{orth}) on $\mathbf{t}$, this sum is well-defined.
Let $T$ denote the set of $\mathbf{t}$ satisfying (\ref{orth}),
which corresponds to the group of characters of $G/K$.
The $\ket{\Psi_{\mathbf{t}}}$ are eigenvectors of each
$U_{f(x \mathbf{e_j})}$, with respective eigenvalues
$\mbox{exp}(2 \pi i x t_j /p^{m_j})$ .
By determining these eigenvalues, for $j=1,2, \ldots, l$,
we will determine
$\mathbf{t}$.
If we had $\ket{\Psi_{\mathbf{t}}}$ in an auxiliary register,
we could estimate $t_j/p^{m_j}$ using $U_{f(\mbox{x}\mathbf{e_j})}$
by the technique of
the previous section.  If we use $F_{p^{m_j}}^{-1}$
we would
determine $t_j$ exactly, or we could use the simpler
$F_{2^k}^{-1}$, for some $k > \log_2(p^{m_j})$, and obtain
$t_j$ with high probability.  For simplicity, we will use
$F_{p^{m_j}}^{-1}$.  In practice we could use $F_{2^k}^{-1}$ for
a large enough $k$ so that the probability of error is
sufficiently small.

By estimating $t_j/p^{m_j}$ for $j = 1,2, \ldots, l$, we determine $\mathbf{t}$.
The algorithm starts by preparing $l$ control registers
in the state $\ket{0}$ and one \emph{target}
or \emph{auxiliary} register in the state $\ket{\Psi_{\mathbf{t}}}$,
applies the appropriate Fourier transforms to produce
\begin{equation}
\left(\sum_{x_1 = 0}^{p^{m_1} - 1} \ket{x_1} \right)
\cdots
\left(\sum_{x_l = 0}^{p^{m_l} - 1} \ket{x_l} \right)
\ket{\Psi_{\mathbf{t}}}
\end{equation}
followed by $U_{f(\mbox{x}{\mathbf{e_j}})}$ for $j=1,2, \ldots, n$,
using the $j$th register as the control and $\ket{\Psi_{\mathbf{t}}}$
as the target, to produce
\begin{equation}
\left(\sum_{x_1 = 0}^{p^{m_1} - 1}
\mbox{exp}(2 \pi i {x_1 t_1 \over p^{m_1}})\ket{x_1 } \right)
\cdots
\left(\sum_{x_l = 0}^{p^{m_l} - 1}
\mbox{exp}(2 \pi i {x_l t_l \over p^{m_l}}) \ket{x_l} \right)
\ket{\Psi_{\mathbf{t}}} .
\end{equation}
Then apply $F_{p^{m_j}}^{-1}$ to the $j$th control register
for each $j$ to yield
\begin{equation}
\ket{t_1} \ket{t_2}
\ldots \ket{t_l}
\ket{\Psi_{\mathbf{t}}}
\end{equation}
from which we can extract $\mathbf{t}$.
As in the previous section, we do not know how to construct
$\ket{\Psi_{\mathbf{t}}}$, but we do know that
\[ \ket{f(\mathbf{0})} =
\sum_{{\mathbf{t}} \in T} \ket{\Psi_{\mathbf{t}}} .\]
So we start with
\[\ket{\mathbf{0}}\ket{\mathbf{0}}\cdots \ket{\mathbf{0}}
\hspace{2mm} \ket{f(\mathbf{0})}
 = \sum_{{\mathbf{t}} \in T} \ket{\mathbf{0}}\ket{\mathbf{0}}\cdots \ket{\mathbf{0}}
\ket{\Psi_{\mathbf{t}}} \]
apply Fourier transforms to get
\begin{equation}
\sum_{{\mathbf{t}} \in T}
\left(\sum_{x_1 = 0}^{p^{m_1} - 1} \ket{x_1} \right)
\cdots
\left(\sum_{x_l = 0}^{p^{m_l} - 1} \ket{x_l} \right)
\ket{\Psi_{\mathbf{t}}}
\label{before}
\end{equation}
then apply $U_{f(\mbox{x}{\mathbf{e_j}})}$
using the $j$th register as a control register, for $j=1,2, \ldots, n$,
and the last register as the target register to produce

\begin{equation}
\sum_{{\mathbf{t}} \in T}
\left(\sum_{x_1 = 0}^{p^{m_1} - 1}
\mbox{exp}(2 \pi i {x_1 t_1 \over p^{m_1}})\ket{x_1} \right)
\cdots
\left(\sum_{x_l = 0}^{p^{m_l} - 1}
\mbox{exp}(2 \pi i {x_l t_l \over p^{m_l}}) \ket{x_l} \right)
\ket{\Psi_{\mathbf{t}}} .
\label{after}
\end{equation}
We finally apply $F_{p^{m_j}}^{-1}$ to the $j$th control register
for $j=1,2, \ldots, l$, to produce
\begin{equation}
 \sum_{{\mathbf{t}} \in T} \ket{\mathbf{t}} \ket{\Psi_t} .
\end{equation}
Observing the first register lets us sample the $\mathbf{t}$'s
uniformly at random, and thus with $O(n)$ repetitions
we will, by (\ref{orth}),
have enough independent linear relations  for us
to determine a generating set for $K$.  For example,
in the case of Simon's problem, the $\ket{\mathbf{t}}$
all satisfy
${\mathbf{t}}\cdot {\mathbf{s}} =
\sum_{j=1}^{l} t_j s_j \mbox{ mod } 2 = 0 \mbox{ mod } 2$,
where $K = \{0,{\mathbf{s}}\}$.
We could also guarantee that each new non-zero element
of $T$ will increase the span by a technique discussed
in the appendix.

This analysis of eigenvectors and eigenvalues is based
on the work in \cite{Kitaev}.
The problem is that, unlike in \cite{Kitaev},
we do not always have the operator $U_{f(\mbox{x}{\mathbf{e_j}})}$.
However, note that, like in Sect. \ref{period.find},
going from (\ref{before}) to (\ref{after})
maps
\[
\left(\sum_{0 \leq x_j \leq p^{m_j}}
\ket{\mathbf{x} } \right)
\ket{f(0)}
\]
 to
\[
\sum_{0 \leq x_j \leq p^{m_j}}
\ket{\mathbf{x}}
\ket{f({\mathbf{x})}} \]

\[ = \sum_{{\mathbf{t}} \in T}
\left(\sum_{x_1 = 0}^{p^{m_1} - 1}
\mbox{exp}(2 \pi i {x_1 t_1 \over p^{m_1}})\ket{x_1} \right)
\cdots
\left(\sum_{x_l = 0}^{p^{m_l} - 1}
\mbox{exp}(2 \pi i {x_l t_l \over p^{m_l}}) \ket{x_l} \right) \ket{\Psi_{\mathbf{t}}}.
\]

We can create state (\ref{after}) by applying $U_f$,
which we do have,
to the starting state
\[  \sum_{0 \leq x_i < p^{m_i}}
\ket{\mathbf{x}} \ket{\mathbf{0}} \]
and proceeding with the remainder of the algorithm.
As in Sect. \ref{period.find}, we are considering the target register
in the basis containing the eigenvectors $\ket{\Psi_k}$ instead
of the computational basis.

\section{Reducing the Size of Control Registers} \label{control.bit}

\subsection{Discrete Logarithms}
In practice, it might be advantageous to
reduce the number of qubits required to solve a problem,
or the length of time each qubit must be isolated from
the environment.
For example, suppose we wish to find $m$ such that
$a^m = b$, where the order of $a$ divides $r$.
The operators $U_{a^x}$ and $U_{b^x}$, which
correspond to multiplication by $a^x$ and $b^x$ respectively,
share the eigenvectors $\ket{\Psi_k}$ (see (\ref{Psi_def_a}))
and have corresponding eigenvalues
$\mbox{exp}(2\pi i kx/r)$ and
$\mbox{exp}(2\pi i kmx/r)$.
We can assume we know $r$ by applying the order-finding algorithm
if necessary.
By using $U_{a^{\mbox{x}}}$ with one control register
we can approximate $k/r$, and by using $U_{b^{\mbox{x}}}$ with
another control register we can approximate $(km \mbox{ mod } r)/r$
and then extract $m$ modulo $r/\mbox{gcd}(r,k)$.  Note that
since we know $r$, we only need
$\log{r}$ bits of precision when estimating $k/r$
and $(km \mbox{ mod } r)/r$, instead of $2 \log_2{r}$ when
using continued fractions.
Note further that, knowing $r$, it may
be possible to actually place
$\ket{\Psi_k}$ into the target register (by direct
construction or otherwise) for some
known $k$,
and thus only require \emph{one} control register with over
$\log_2 r$ qubits to estimate $(km \mbox{ mod } r)/r$.
One way of doing this is to keep the target
register after we have applied the order-finding
algorithm and observed an estimate of $k/r$ in
the control register.  At this point,
the target register is almost entirely
in the state $\ket{\Psi_k}$, and we
could now just estimate
the eigenvalue of $U_{b^x}$ on this eigenstate, which
we know will be $(km  \mbox{ mod } r)/r$.

\subsection{One Control Bit}

Consider the case that we have an efficient computational
means of mapping
$\ket{f(\mathbf{y})}$ to $\ket{f({\mathbf{y}} + {\mathbf{x}} )}$
for any $\mathbf{x}$.
If we consider
$f$ to be of the form $h\circ g$ for a homomorphism $g$,
we are requiring that $h$ is the identity or some other function
with enough structure that we can efficiently map
$h(g({\mathbf{y}}))$ to
$h(g({\mathbf{y}}+{\mathbf{x}})) = h(g({\mathbf{y}})+g({\mathbf{x}}))$.
In this case we can efficiently solve the hidden subgroup
problem with only one control bit or a sequence of
\emph{flying qubits} \cite{THLMK}.
We illustrate this method for the problem of
finding the order of an element $a$ in a group $H$.

\begin{figure}
\psfig{figure=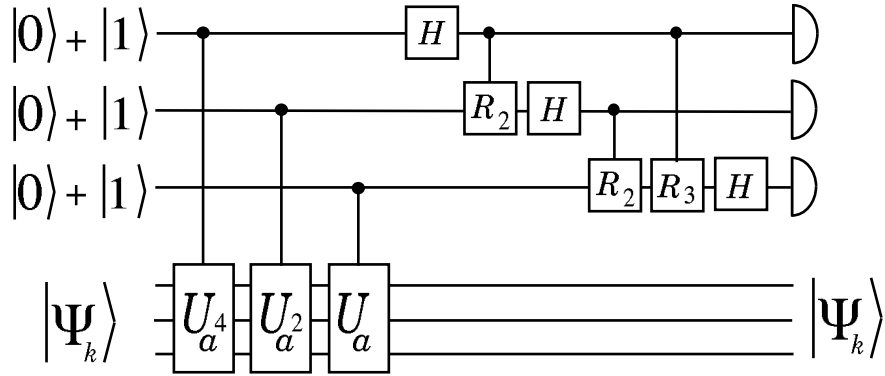}
\caption[ ]{We start with
$(\ket{0} + \ket{1})(\ket{0} + \ket{1})(\ket{0} + \ket{1}) \ket{\Psi_k}$
$= \sum_{x=0}^{7} \ket{x} \ket{\Psi_k}$.  The controlled
multiplications create the state
$\sum_{x=0}^{7} \mbox{exp}(2 \pi i k/r) \ket{x} \ket{\Psi_k}$.
The remaining gates create the state $\sket{\widetilde{k/r}}$
(apart from reversing the order of the qubits) which we then observe.
The $H$-gates correspond to \emph{Hadamard} transforms,
and the $R_j$-gates correspond to a controlled phase shift of
$\mbox{exp}(2\pi i/2^{j})$ on state $\ket{1}$.}
\label{figure.1}
\end{figure}

\begin{figure}
\psfig{figure=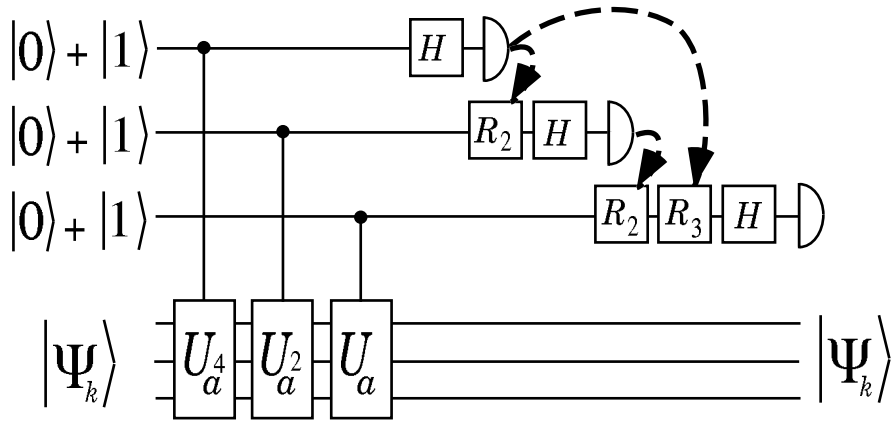}
\caption[ ]{Here we employ a `semi-classical' version of $F_{2^3}^{-1}$.
We could measure each qubit before it is used as a control, perform
the controlled rotations `semi-classically', and the probability of
observing each possible output state $\ket{x_1}\ket{x_2}\ket{x_3}$
is the same as in Fig. \ref{figure.1}. }
\label{figure.2}
\end{figure}

Figure \ref{figure.1} shows the relationship between $F_{2^n}^{-1}$
and the controlled multiplications by powers of $a$ in the
order-finding algorithm. As already pointed out in
\cite{Griffiths.Niu}, the measurements could be performed before
the controlled rotations. The quantum controlled rotations could
then be replaced with `semi-classically' controlled rotations of
the subsequent qubits (that is, the control bit is measured and, if
the outcome is $1$, the rotation is done quantumly). This brings us
to Fig. \ref{figure.2}, where we observe further that all the
operations on the first qubit could be performed before we even
prepare the second qubit.  All the operations could be done
sequentially, starting from the first qubit, the results of
measuring the previous qubits determining how to prepare the next
qubit before measurement. This means we could in fact do all the
quantum controlled multiplications with a single control qubit
provided we can execute the `semi-classical' controls which allow
us to reset a qubit to $\ket{0} + \ket{1}$ and perform a rotation
dependent upon the previous measurements (the rotations could in
fact be implemented at any time after resetting the qubit and
before applying the final Hadamard transform and measuring it; they
could also be omitted provided we repeat each step a few extra
times and do some additional classical post-processing as done in
\cite{Kitaev}). Alternatively, the control qubits could be a
sequence of flying qubits which are measured (or prepared) in a way
dependent upon the outcomes of the previous measurements of control
qubits.

For the more general hidden subgroup problem in Abelian groups we
would have a sequence of applications of
$U_{f(\mbox{x}{\mathbf{e_j}})}$ controlled by one qubit, which is
measured, then reset to a superposition of $\ket{0}$ and $\ket{1}$
plus some rotation that is dependent upon the previous
measurements. In summary:

\vspace{3mm}
\noindent
\emph{
The hidden subgroup $K$ of a finitely generated Abelian group $G$
generated by ${\mathbf{e_1}},{\mathbf{e_2}}, \ldots
{\mathbf{e_k}}$, corresponding to a function $f$ from $G$ to a
finite set $X$, can be found with probability close to $1$ by
`semi-classical' methods with only \emph{one} control bit (or a
sequence of flying qubits) and polynomial in $n$ applications of
the operators $\ket{x}\ket{f({\mathbf{y}})} \rightarrow
\ket{x} \ket{f({\mathbf{y}} + x{\mathbf{e_j}})}$
for $j = 1, 2, \ldots, k$, where $n$ is the index of $K$ in $G$.}

\section*{Acknowledgments}
Many thanks to Peter H{\o}yer for helping prepare this paper,
to Mark Ettinger and Richard Hughes for helpful
discussions and hospitality in Los Alamos,
to BRICS (Basic Research in Computer Science, Centre of the
Danish National Research Foundation), and to
Wolfson College.

  This work was supported in part by CESG,
  the European TMR Research Network ERP-4061PL95-1412,
  Hewlett-Packard, The Royal Society London,
  and the U.S. National Science Foundation under Grant No. PHY94-07194.
  Part of this work was done at the 1997 Elsag-Bailey -- I.S.I. Foundation
  workshop on quantum computation, at NIS-8 division of the
  Los Alamos National Laboratory, and at the BRICS 1998 workshop on
  Algorithms in Quantum Information Processing.

\section*{Appendix: When $f$ is many-to-1 on $G/K$} \label{m_to_1}

The question of what happens when $f$ is many-to-1 on
cosets of $K$
was first addressed in \cite{Boneh-Lipton}.
This is a slight weakening of the promise that $f$ is
distinct on each coset.
Suppose $f$ can have up to $m$ cosets going to
the same output, for some known $m$.  That is,
$f = h \circ g$ where $g$ is a homomorphism from $G$ to a some group $H$
with kernel $K$, and $h$ is a mapping from $H$ to $X$ that
is at most $m$-to-1.
If $m$ divides the order of $K$,
we clearly have a problem.  For example, suppose
$K$ is the cyclic group of order $2M$, and
$m = 2$, but by changing one value of $f$ it would have
period $M$.  It can easily be shown that
$\Omega(\sqrt{M})$ (that is, at least $c \sqrt{M}$ for
some positive constant $c$) applications of
$f$ are necessary to distinguish such a modified $f$
from the original one with probability greater than $3/4$,
and thus
no polynomial time algorithm, quantum or classical,
could distinguish the two cases.
Thus one requirement for there to exist an efficient solution
in the worst case is that $m$ is  less than
the smallest prime factor of $|K|$, the number of elements
in $K$.

The problem when $f$ is not 1-to-1 is the following.
Running the same quantum algorithm will produce
the state
\[ \sum_{k=0}^{r-1} \sket{\widetilde{{k/r}}} \ket{\Psi_k^{\prime}} \]
where
\[ \ket{\Psi_k^{\prime}}
= \sum_{t = 0}^{r-1} \mbox{exp}(-2\pi i kt/r)\ket{f(t)}.\]
This is the same definition as in
(\ref{Psi_def_gen}) except now the $\ket{f(t)}$
are not necessarily distinct.  This means the sizes of
each of the $\ket{\Psi_k^{\prime}}$  are not
necessarily the same since both destructive and constructive
interference can occur.
Also, the $\ket{\Psi_k^{\prime}}$
are no longer orthogonal, and thus some constructive
interference could occur on the poor estimates of
$k/r$.
Recall that even the close estimates of $k/r$
will not yield useful results
when $k = 0$. Any other $k$ will at least
reveal a small factor of $r$.
So we need to guarantee that the probability of
observing a close enough estimate of
$ k/r $  for some $k \neq 0$
is significant.

By making our estimates precise enough, say
by using  over $2\log_2 r + \epsilon/m^2 $ control
qubits, the estimates of $k/r$ will have error less than
$1/2r^2$ (so that continued fractions will work)
with probability at least
$1 - \epsilon/m^2$.  Thus assuming $f$ is 1-to-1,
the probability of observing a \emph{bad}
output other than $0$
would be at most $\epsilon/m^2$, and the probability
of observing $0$ would be at most $1/r + \epsilon/m^2$.
However, since $f$ is at most $m$-to-1, these probabilities could
amplify by at most a factor of $m^2$
to $\epsilon$ and $m^2/r + \epsilon$ respectively.
Observing a $0$ means we either got a bad
output, or the period of $f$ is $1$.
Getting $0$ as a bad output is not very harmful,
however getting another bad output is more complicated,
since it will give us a false factor of $r$.
It will be useful to make $\epsilon$ small, so that
it is unlikely our answer is tainted by false factors of $r$.
Once we have one factor $r_1$ of $r$, we can
replace $f(x)$ with $f(r_1 x)$
(as done in \cite{Boneh-Lipton}),
which has period $r/r_1$
and find a
factor of $r/r_1$.
Once we have a big enough factor $r^{\prime}$ of $r$,
we might start observing $0$'s, which
tells us that  the remaining factor of the original
$r$, namely $r/r^{\prime}$, is less than $m^2$.
Thus we can explicitly test
$f(r^{\prime}), f(2r^{\prime}), f(3r^{\prime}), \ldots, $
until we find the period, which will occur
after at most $m^2$ applications.
We thus have an algorithm with running time, in terms
of elementary quantum operations and applications of $f$,
polynomial in $\log(r)$ and quadratic in $m$.

The trick of reducing the order of the function
can be applied to reduce the size of the group
and hidden subgroup in the finite Abelian
hidden subgroup problem.
When $G = {\mathbf{Z}}_p$, we can efficiently test if $K = G$ or $K = \{1\}$.
The above analysis tells us how to deal with the
case that $G = {\mathbf{Z}}_{p^l}$ for $n>1$.  A similar technique
will reduce
$G = {\mathbf{Z}}_{p^{l_1}} \times \cdots {\mathbf{Z}}_{p^{l_k}}$
to a quotient group $\overline{G}$ and we can again proceed
inductively until the size of $\overline{G}$ is less than
$m^2$.  We can then exhaustively test $\overline{G}$
for the hidden subgroup $\overline{K}$ in another $O(m^2)$ steps.

We emphasize that this is a worst-case analysis.  If there
were a noticeable difference in the behaviour of a 1-to-1
and an $m$-to-1 function $f$, $m>1$, we could decide if a given function
$h$ is 1-to-1 or many-to-one (by composing $h$ with a function $f$
whose period or hidden Abelian subgroup we know, and
test for this difference in behaviour).
Distinguishing 1-to-1 functions from many-to-1 functions
seems like a \emph{very}
difficult task in general, and would solve
the graph automorphism problem, for example.

\end{document}